\documentclass[reqno,11pt]{amsart}
\usepackage[utf8]{inputenc}
\usepackage{graphicx}
\usepackage{slashed}
\usepackage{amscd}
\usepackage{tensor}
\usepackage{amssymb}
\usepackage{esint}
\usepackage{enumitem}
\usepackage[off]{auto-pst-pdf} 
\usepackage[mathscr]{eucal}
\numberwithin{equation}{section}
\allowdisplaybreaks[1]

\title[Global Solutions to Linearized Field Equations]{Construction of Global Solutions to the Linearized
Field Equations for Causal Variational Principles}

\author[F.\ Finster]{Felix Finster}
\address{Fakult\"at f\"ur Mathematik \\ Universit\"at Regensburg \\ D-93040 Regensburg \\ Germany}
\email{finster@ur.de}

\author[M.\ Kraus]{Margarita Kraus \\ \\ October 2022}
\address{Fachbereich 08 \\ Johannes-Gutenberg-Univer\-si\-t\"at Mainz \\
D-55099 Mainz \\ Germany }
\email{mkraus@mathematik.uni-mainz.de}

\newtheorem{Def}{Definition}[section]
\newtheorem{Thm}[Def]{Theorem}

\newcommand{\Thanks}{\vspace*{.5em} \noindent \thanks}
\newcommand{\beq}{\begin{equation}}
\newcommand{\eeq}{\end{equation}}
\newcommand{\Proof}{\begin{proof}}
\newcommand{\QED}{\end{proof} \noindent}

\newcommand{\la}{\langle}
\newcommand{\ra}{\rangle}

\newcommand{\R}{\mathbb{R}}

\newcommand{\N}{\mathbb{N}}

\renewcommand{\L}{{\mathcal{L}}}
\newcommand{\Sact}{{\mathcal{S}}}

\DeclareMathOperator{\supp}{supp}

\newcommand{\F}{{\mathscr{F}}}

\newcommand{\D}{\mathscr{D}}

\newcommand{\itemD}{\item[{\raisebox{0.125em}{\tiny $\blacktriangleright$}}]}

\DeclareFontFamily{OT1}{rsfso}{}
\DeclareFontShape{OT1}{rsfso}{m}{n}{ <-7> rsfso5 <7-10> rsfso7 <10-> rsfso10}{}
\DeclareMathAlphabet{\mycal}{OT1}{rsfso}{m}{n}

\newcommand{\J}{\mathfrak{J}}
\newcommand{\s}{\mathfrak{s}}
\newcommand{\Jdiff}{\mathfrak{J}^\text{\rm{\tiny{diff}}}}
\newcommand{\Jtest}{\mathfrak{J}^\text{\rm{\tiny{test}}}}
\newcommand{\Jlin}{\mathfrak{J}^\text{\rm{\tiny{lin}}}}
\newcommand{\Jvary}{\mathfrak{J}^\text{\rm{\tiny{vary}}}}
\renewcommand{\u}{\mathfrak{u}}
\renewcommand{\v}{\mathfrak{v}}
\newcommand{\w}{\mathfrak{w}}
\newcommand{\Gdiff}{\Gamma^\text{\rm{\tiny{diff}}}}
\newcommand{\Gtest}{\Gamma^\text{\rm{\tiny{test}}}}

\newcommand{\bitem}{\begin{itemize}[leftmargin=2em]}
\newcommand{\eitem}{\end{itemize}}

\newcommand{\tmax}{{t_{\max}}}
\newcommand{\tmin}{{t_{\min}}}

\newcommand{\vary}{\text{\rm{\tiny{vary}}}}

\newcommand{\loc}{\text{\rm{loc}}}

\newcommand{\Ctest}{C^\text{\rm{\tiny{test}}}}
\renewcommand{\sc}{\text{\rm{sc}}}

\begin{document}

\maketitle

\begin{abstract}
We give a novel construction of global solutions to the linearized field equations for
causal variational principles. The method is to glue together local solutions 
supported in lens-shaped regions. As applications, causal Green's operators 
and cone structures are introduced.
\end{abstract}

\tableofcontents

\section{Introduction} \label{intro}

Causal variational principles were introduced in~\cite{continuum}
as a mathematical generalization of the causal action principle,
being the analytical core of the physical theory of causal fermion systems
(for the general context see the reviews~\cite{ nrstg, dice2014}, the textbooks~\cite{cfs, intro}
or the website~\cite{cfsweblink}).
In general terms, given a manifold~$\F$ together with a non-negative function~$\L : \F \times \F \rightarrow \R^+_0$,
in a {\em{causal variational principle}} one minimizes the action~$\Sact$ given by
\[ 
\Sact (\rho) = \int_\F d\rho(x) \int_\F d\rho(y)\: \L(x,y) \]
under variations of the measure~$\rho$ on~$\F$, keeping the total volume~$\rho(\F)$ fixed
(for the precise mathematical setup see Section~\ref{seccvp} below).
The support of the measure~$\rho$ denoted by
\beq \label{Mdef}
M := \supp \rho \subset \F
\eeq
has the interpretation as the underlying {\em{space}} or {\em{spacetime}}.
A minimizing measure satisfies corresponding {\em{Euler-Lagrange (EL) equations}}
(for details see the preliminaries in Section~\ref{seccvp}
or~\cite[Chapter~7]{intro}).
For the detailed analysis of minimizing measures, it is very useful to consider first variations of the measure~$\rho$
which preserve the EL equations. Such a variation is described by 
a so-called {\em{jet}}~$\v=(b,v)$, which consists of a scalar function~$b$ and a vector field~$v$
(for details see~\eqref{Jdef} in Section~\ref{secEL} below). Moreover, this jet satisfies
the {\em{homogeneous linearized field equations}}
\[ \Delta \v = 0 \:, \]
where the operator~$\Delta$ is defined by
\beq \label{Deltadef}
\Delta \v(x) = \nabla \bigg( \int_M \big( \nabla_{1, \v} + \nabla_{2, \v} \big) \L(x,y)\: d\rho(y) - \nabla_\v \,\s \bigg) \:.
\eeq
Here~$\s$ is a positive parameter, and the jet derivative~$\nabla$ is a combination of
multiplication and directional derivative (for details see again Section~\ref{secEL}).
For the mathematical analysis of the linearized field equations, it is preferable to
include an inhomogeneity~$\w$,
\beq \label{Delvw}
\Delta \v = \w \:.
\eeq
An introduction to the linearized equations and more details can be found in~\cite[Chapter~14]{intro}.

In~\cite{linhyp} the notion of weak solution was introduced, and energy methods were used to
construct weak solutions locally in so-called lens-shaped regions.
Moreover, global solutions were constructed under the assumptions that spacetime can be
exhausted by lens-shaped regions, and that so-called {\em{shielding conditions}} hold.
In~\cite{weyl}, on the other hand, the existence of global foliations by surface layers
was assumed.
Although all these assumptions are physically sensible, they seem too strong for some applications.
First, exhaustions by lens-shaped regions exclude some cases of non-trivial
spacetime topology. Likewise, assuming the existence of global foliations
seems too strong, because it is unclear whether a
statement similar to the existence of global smooth foliations in the globally hyperbolic setting in~\cite{bernal+sanchez} holds for causal variational principles.
Second and more importantly, the shielding conditions have the disadvantage that they
are quite strong and difficult to verify in the applications. This raises the important question how to
construct global solutions without referring to exhaustions by lens-shaped regions and to shielding conditions.
In the present paper we answer this question by giving an alternative construction of global solutions
which applies under general assumptions without using shielding conditions
(see Theorem~\ref{thmglobal}). Our construction also has the advantage that it is considerably simpler,
thereby also clarifying the resulting causal structure of spacetime.

The basic idea underlying our construction is to ``glue together local solutions'' supported in
lens-shaped regions. Such gluing constructions are familiar from the theory of hyperbolic
PDEs (see~\cite{friedrichs}, the textbooks~\cite{courant+hilbert2, john, taylor3}, \cite[Chapter~13]{intro}
or similarly in globally hyperbolic spacetimes~\cite{rendall, ringstroem, baergreen}).
In this setting, the uniqueness of solutions of the Cauchy problem implies that the local solutions
coincide on the intersection of their domains, making it possible to obtain the global solution simply
by demanding that its restriction to every lens-shaped region coincides with the respective local solution
(for details see for example~\cite[Section~13.3]{intro}).
This method does not apply in the setting of causal variational principles, because the
linearized field operator~\eqref{Deltadef} is non-local. Therefore, ``localizing'' a solution 
by multiplying with cutoff functions gives rise to
error terms, which need to be controlled. Another complication is that, in the weak formulation of the
equations, the space used for testing need not be dense. Therefore, solutions are unique only up to
vectors in the orthogonal complement of the test space. 
In particular, two solutions constructed in two lens-shaped regions will in general not coincide
on the intersection of the lens-shaped regions. Therefore, it is a non-trivial task to extend the
solution to the union of the two lens-shaped regions.
These difficulties must be taken into account when ``gluing together'' local solutions.

In order to overcome these difficulties, we proceed as follows.
We begin with local solutions defined in lens-shaped regions. By multiplying with a cutoff function
near the future boundary and extending to zero, we obtain a global solution, however with an
additional inhomogeneity near the future boundary. This construction also makes it necessary to
adjust the space of admissible test functions. 
As will be arranged and made precise in our construction, the support of the additional inhomogeneity
lies in the future of the original inhomogeneity. This allows us to proceed inductively,
to the effect that, visually speaking, the support of the inhomogeneity ``moves to future infinity''
(see Figure~\ref{figfuturerel} on page~\pageref{figfuturerel} and the precise statements in Section~\ref{secglue}).

The paper is organized as follows. Section~\ref{secprelim} provides the necessary preliminaries
on causal variational principles and the linearized field equations.
In Section~\ref{secglobal} global solutions are constructed.
In Section~\ref{segreen} we proceed by constructing
causal Green's operators~$S^\wedge$ and~$S^\vee$ as well as the causal fundamental solution~$G$.
We show that the properties of the operators~$\Delta$ and~$G$ can be summarized in the
exact sequence
\beq \label{exact}
0 \rightarrow \Jtest_0 \overset{\Delta}{\longrightarrow} \J^*_0
\overset{G}{\longrightarrow} \J_\sc
\overset{\Delta}{\longrightarrow} \J^*_\sc \rightarrow 0 \:,
\eeq
where~$\Jtest_0$ and~$\J_0^*$ are suitable spaces of compactly supported jets, whereas~$\J_\sc$
and~$\J^*_\sc$ have spatially compact support (see Theorem~\ref{thmexact}).
In Section~\ref{seccones} we conclude the paper with a suggestion on how to construct
causal cones and transitive causal structures.

\section{Preliminaries} \label{secprelim}

\subsection{Causal Variational Principles in the Non-Compact Setting} \label{seccvp}
We consider causal variational principles in the non-compact setting as
introduced in~\cite[Section~2]{jet}. Thus we let~$\F$ be a (possibly non-compact)
smooth manifold of dimension~$m \geq 1$
and~$\rho$ a (positive) Borel measure on~$\F$.
Moreover, the
\[ \text{\em{Lagrangian}} \qquad \L : \F \times \F \rightarrow \R^+_0 \]
is defined as a non-negative function with the following properties:
\bitem
\item[(i)] $\L$ is symmetric: $\L(x,y) = \L(y,x)$ for all~$x,y \in \F$.\label{Cond1}
\item[(ii)] $\L$ is strictly positive on the diagonal: $\L(x,x) >0$ for all~$x \in \F$.
\item[(iii)] $\L$ is lower semi-continuous, i.e.\ for all sequences~$x_n \rightarrow x$ and~$y_{n'} \rightarrow y$,
\[ \L(x,y) \leq \liminf_{n,n' \rightarrow \infty} \L(x_n, y_{n'})\:. \]\label{Cond2}
\eitem
The {\em{causal variational principle}} is to minimize the
\beq \label{Sact} 
\text{\em{action}} \qquad \Sact (\rho) = \int_\F d\rho(x) \int_\F d\rho(y)\: \L(x,y) 
\eeq
under variations of the measure~$\rho$, keeping the total volume~$\rho(\F)$ fixed
({\em{volume constraint}}). According to~\eqref{Mdef},
{\em{spacetime}}~$M$ is defined as the support of the measure~$\rho$.
It is a (not necessarily smooth) subset of~$\F$.
The notion {\em{causal}} in ``causal variational principles'' refers to the fact that
the Lagrangian induces on~$M$ a {\em{causal structure}} given by
\[ 
\text{$x,y \in M$ are } \left\{ \begin{array}{c} \text{timelike} \\ \text{spacelike} \end{array} \right\}
\text{ separated if } \left\{ \begin{array}{c} \L(x,y)>0 \\ \L(x,y)=0 \end{array} \right\} \:. \]
An important example of causal variational principles is the {\em{causal action principle}} for
{\em{causal fermion systems}} (for the connection see~\cite[Section~2]{jet}).

If the total volume~$\rho(\F)$ is finite, one minimizes~\eqref{Sact}
over all regular Borel measures with the same total volume.
If the total volume~$\rho(\F)$ is infinite, however, it is not obvious how to implement the volume constraint,
making it necessary to proceed as follows.
We need the following additional assumptions:
\bitem
\item[(iv)] The measure~$\rho$ is {\em{locally finite}}
(meaning that any~$x \in \F$ has an open neighborhood~$U$ with~$\rho(U)< \infty$).\label{Cond3}
\item[(v)] The function~$\L(x,.)$ is $\rho$-integrable for all~$x \in \F$, and integration gives
a lower semi-continuous and bounded function on~$\F$. \label{Cond4}
\eitem
Given a regular Borel measure~$\rho$ on~$\F$, we can vary over all
regular Borel measures~$\tilde{\rho}$ with
\[ 
\big| \tilde{\rho} - \rho \big|(\F) < \infty \qquad \text{and} \qquad
\big( \tilde{\rho} - \rho \big) (\F) = 0 \]
(where~$|.|$ denotes the total variation of a measure).
The existence theory for minimizers is developed in~\cite{noncompact}.

\subsection{The Euler-Lagrange Equations and Jet Spaces} \label{secEL}
A minimizer of the causal variational principle
satisfies the following {\em{Euler-Lagrange (EL) equations}}:
For a suitable value of the parameter~$\s>0$,
the lower semi-continuous function~$\ell : \F \rightarrow \R_0^+$ defined by
\[ 
\ell(x) := \int_\F \L(x,y)\: d\rho(y) - \s \]
is minimal and vanishes on spacetime~$M:= \supp \rho$,
\beq \label{EL}
\ell|_M \equiv \inf_\F \ell = 0 \:.
\eeq
The parameter~$\s$ can be understood as the Lagrange parameter
corresponding to the volume constraint. For the derivation and further details we refer to~\cite[Section~2]{jet}
or~\cite{noncompact}.

The EL equations~\eqref{EL} are nonlocal in the sense that
they make a statement on the function~$\ell$ even for points~$x \in \F$ which
are far away from spacetime~$M$.
It turns out that for the applications we have in mind, it is preferable to
evaluate the EL equations only locally in a neighborhood of~$M$.
This leads to the {\em{restricted EL equations}} introduced in~\cite[Section~4]{jet}.
Here we give a slightly less general version of these equations which
is sufficient for our purposes. In order to explain how the restricted EL equations come about,
we begin with the simplified situation that the function~$\ell$ is smooth.
In this case, the minimality of~$\ell$ implies that the derivative of~$\ell$
vanishes on~$M$, i.e.\
\beq \label{ELweak}
\ell|_M \equiv 0 \qquad \text{and} \qquad D \ell|_M \equiv 0
\eeq
(where~$D \ell(p) : T_p \F \rightarrow \R$ is the derivative).
In order to combine these two equations in a compact form,
it is convenient to consider a pair~$\u := (a, u)$
consisting of a real-valued function~$a$ on~$M$ and a vector field~$u$
along the embedding~$M \hookrightarrow \F$, and to denote the combination of 
multiplication and directional derivative by
\beq \label{Djet}
\nabla_{\u} \ell(x) := a(x)\, \ell(x) + \big(D_u \ell \big)(x) \:.
\eeq
Then the equations~\eqref{ELweak} imply that~$\nabla_{\u} \ell(x)$
vanishes for all~$x \in M$ and for all pairs~$(a,u)$.
The pair~$\u=(a,u)$ is referred to as a {\em{jet}}.

In the general lower-continuous setting, one must be careful because
the directional derivative~$D_u \ell$ in~\eqref{Djet} need not exist.
Our method for dealing with this problem is to restrict attention to vector fields
for which the directional derivative is well-defined.
Moreover, we must specify the regularity assumptions on~$a$ and~$u$.
To begin with, we always assume that~$a$ and~$u$ are {\em{smooth}} in the sense that they
belong to the jet space
\beq \label{Jdef}
\J := \big\{ \u = (a,u) \text{ with } a \in C^\infty(M, \R) \text{ and } u \in \Gamma(M, T\F) \big\} \:,
\eeq
where~$C^\infty(M, \R)$ and~$\Gamma(M,T\F)$ denote the space of real-valued functions and vector fields
on~$M$, respectively, which admit a smooth extension to~$\F$.

Clearly, the fact that a jet~$\u$ is smooth does not imply that the functions~$\ell$
or~$\L$ are differentiable in the direction of~$\u$. This must be ensured by additional
conditions which are satisfied by suitable subspaces of~$\J$
which we now introduce.
First, we let~$\Gdiff$ be those vector fields for which the
directional derivative of the function~$\ell$ exists,
\[ \Gdiff = \big\{ u \in C^\infty(M, T\F) \;\big|\; \text{$D_{u} \ell(x)$ exists for all~$x \in M$} \big\} \]
(in typical applications, this space is non-zero, as one sees in the example of the smooth compact setting in~\cite[Section~6.2]{intro}).
This gives rise to the jet space
\[ \Jdiff := C^\infty(M, \R) \oplus \Gdiff \;\subset\; \J \:. \]
For the jets in~$\Jdiff$, the combination of multiplication and directional derivative
in~\eqref{Djet} is well-defined. 
Next, we choose a linear subspace~$\Jtest \subset \Jdiff$ with the property
that its scalar and vector components are both vector spaces, i.e.\
\[ \Jtest = \Ctest(M, \R) \oplus \Gtest \;\subseteq\; \Jdiff \:, \]
with subspaces~$\Ctest(M, \R) \subset C^\infty(M, \R)$ and~$\Gtest \subset \Gdiff$.
We assume that the scalar component is nowhere trivial in the sense that
\[ 
\text{for all~$x \in M$ there is~$a \in \Ctest(M, \R)$ with~$a(x) \neq 0$}\:. \]
The {\em{restricted EL equations}} read (for details cf.~\cite[(eq.~(4.10)]{jet})
\beq \label{ELtest}
\nabla_{\u} \ell|_M = 0 \qquad \text{for all~$\u \in \Jtest$}\:.
\eeq
The purpose of introducing~$\Jtest$ is that it gives the freedom to restrict attention to the portion of
information in the EL equations which is relevant for the application in mind.

Before going on, we point out that the restricted EL equations~\eqref{ELtest}
do not hold only for minimizers, but also for critical points of
the causal action. With this in mind, all methods and results of this paper 
do not apply only to
minimizers, but more generally to critical points of the causal variational principle.
For brevity, we also refer to a measure with satisfies the restricted EL equations~\eqref{ELtest}
as a {\em{critical measure}}.

We conclude this section by introducing a few other jet spaces which will be needed
later on. It is useful to define the differentiability properties of the jets by corresponding
differentiability properties of the Lagrangian.
When considering higher derivatives, we always choose 
charts and work in components.
For ease in notation, we usually omit all vector and tensor indices.
But one should keep in mind that, from now on, we always work in suitably chosen charts.
Here and throughout this paper, we use the following conventions for partial derivatives and jet derivatives:
\bitem
\itemD Partial and jet derivatives with an index $i \in \{ 1,2 \}$, as for example in~\eqref{derex}, only act on the respective variable of the function $\L$.
This implies, for example, that the derivatives commute,
\[ 
\nabla_{1,\v} \nabla_{1,\u} \L(x,y) = \nabla_{1,\u} \nabla_{1,\v} \L(x,y) \:. \]
\itemD The partial or jet derivatives which do not carry an index act as partial derivatives
on the corresponding argument of the Lagrangian. This implies, for example, that
\[ \nabla_\u \int_\F \nabla_{1,\v} \, \L(x,y) \: d\rho(y) =  \int_\F \nabla_{1,\u} \nabla_{1,\v}\, \L(x,y) \: d\rho(y) \:. \]
\eitem
We point out that, in contrast to the method and conventions used in~\cite{jet},
{\em{jets are never differentiated}}.

We next introduce the jet spaces~$\J^\ell$, where~$\ell \in \N \cup \{\infty\}$ can be
thought of as the order of differentiability if the derivatives act simultaneously on
both arguments of the Lagrangian:
\begin{Def} \label{defJvary}
For any~$\ell \in \N_0  \cup \{\infty\}$, the jet space~$\J^\ell \subset \J$
is defined as the vector space of jets with the following properties:
\begin{itemize}[leftmargin=2.5em]
\item[\rm{(i)}] For all~$y \in M$ and all~$x$ in an open neighborhood of~$M$,
the directional derivatives
\beq \label{derex}
\big( \nabla_{1, \v_1} + \nabla_{2, \v_1} \big) \cdots \big( \nabla_{1, \v_p} + \nabla_{2, \v_p} \big) \L(x,y)
\eeq
(computed componentwise in charts around~$x$ and~$y$)
exist for all~$p \in \{1, \ldots, \ell\}$ and all~$\v_1, \ldots, \v_p \in \J^\ell$. Here the subscripts $1,2$ refer to the derivatives acting on the first and on the second argument of $\L(x,y)$ respectively.
\item[\rm{(ii)}] The functions in~\eqref{derex} are $\rho$-integrable
in the variable~$y$, giving rise to locally bounded functions in~$x$. More precisely,
these functions are in the space
\[ L^\infty_\text{\rm{loc}}\Big( M, L^1\big(M, d\rho(y) \big); d\rho(x) \Big) \:. \]
\item[\rm{(iii)}] Integrating the expression~\eqref{derex} in~$y$ over~$M$
with respect to the measure~$\rho$,
the resulting function (defined for all~$x$ in an open neighborhood of~$M$)
is continuously differentiable in the direction of every jet~$\u \in \Jtest$.
\eitem
\end{Def} \noindent

Finally, compactly supported jets are denoted by a subscript zero, like for example
\beq \label{J0def}
\Jtest_0 := \{ \u \in \Jtest \:|\: \text{$\u$ has compact support} \} \:.
\eeq

\subsection{The Linearized Field Equations} \label{seclinfield}
In words, the {\em{homogeneous linearized field equations}}
describe variations of the measure~$\rho$ which preserve the EL equations.
In order to make this statement mathematically precise,
we consider variations where we multiply~$\rho$ by a non-negative
function and take the push-forward with respect to a mapping from~$M$ to~$\F$.
Thus we consider families of measures~$(\tilde{\rho}_\tau)_{\tau \in (-\delta, \delta)}$ 
of the form
\beq \label{rhotau}
\tilde{\rho}_\tau = (F_\tau)_* \big( f_\tau \, \rho \big) \:,
\eeq
where~$f$ and~$F$ are smooth,
\[ f \in C^\infty\big((-\delta, \delta) \times M, \R^+ \big) \qquad \text{and} \qquad
F \in C^\infty\big((-\delta, \delta) \times M, \F \big) \:, \]
and have the properties~$f_0(x)=1$ and~$F_0(x) = x$ for all~$x \in M$
(here the push-forward measure is defined
for a subset~$\Omega \subset \F$ by~$((F_\tau)_*\rho)(\Omega)
= \rho ( F_\tau^{-1} (\Omega))$; see for example~\cite[Section~3.6]{bogachev}).
Demanding that the family of measures~\eqref{rhotau} is critical for all~$\tau$
implies that the jet~$\v$ defined by
\[ 
\v(x) := \frac{d}{d\tau} \big( f_\tau(x), F_\tau(x) \big) \Big|_{\tau=0} \]
satisfies the linearized field equations
\beq \label{Lapdef}
0 = \la \u, \Delta \v \ra(x) := 
\nabla_\u \bigg( \int_M \big( \nabla_{1, \v} + \nabla_{2, \v} \big) \L(x,y)\: d\rho(y) - \nabla_\v \,\s \bigg)
\eeq
for all~$\u \in \Jtest$ and all~$x \in M$
(for the derivation see~\cite[Section~3.3]{perturb}
and~\cite[Section~4.2]{jet}).
Since these equations hold pointwise in~$x$, we
here refer to these equations as the {\em{strong}} equations
(in distinction of the weak equations obtained by testing and integrating; see
Section~\ref{secweak} below).

The {\em{inhomogeneous}} linearized field equations are obtained by adding an inhomogeneity
on the right side of the homogeneous equations~\eqref{Lapdef}, i.e.\
\beq \label{lininhom}
\la \u, \Delta \v \ra|_M = \la \u, \w \ra \qquad \text{for all~$\u \in \Jtest$} \:.
\eeq
One way to give the right side of this equation a precise meaning is to
regard~$\w$ as a dual jet, so that~$\la \u, \w \ra$ is a dual pairing
(for details see~\cite[Sections~2.2. and~2.3]{linhyp}). In what follows, it is more suitable to identify jets and dual
jets by a scalar product (for details see~\cite[Section~3.2]{linhyp}). To this end,
we let~$\Gamma_x$ be the subspace of the tangent space spanned by the test jets,
\[ 
\Gamma_x := \big\{ u(x) \:|\: u \in \Gtest \big\} \;\subset\; T_x\F\:. \]
We introduce a Riemannian metric~$g_x$ on~$\J_x$.
This Riemannian metric also induces a pointwise scalar product on the jets. Namely, setting
\[ 
\J_x := \R \oplus \Gamma_x \:, \]
we obtain the scalar product on~$\J_x$
\beq
\la \v, \tilde{\v} \ra_x \,:\, \J_x \times \J_x \rightarrow \R \:,\qquad
\la \v, \tilde{\v} \ra_x := b(x)\, \tilde{b}(x) + g_x \big(v(x),\tilde{v}(x) \big) \:. \label{vsprod}
\eeq
Using this scalar product,
the inhomogeneity in~\eqref{lininhom} is well-defined for any jet~$\w \in \Jtest$.
We point out that the inhomogeneous equation~\eqref{lininhom}
is again evaluated pointwise for every~$x \in M$. Therefore, we
refer to it as the {\em{strong}} linearized field equations.
For brevity, sometimes we leave out the pointwise testing and write 
this equation in the shorter form~\eqref{Delvw}.

\subsection{Surface Layer Integrals} \label{secosi}
{\em{Surface layer integrals}} were first introduced in~\cite{noether}
as double integrals of the general form
\beq \label{osi}
\int_\Omega \bigg( \int_{M \setminus \Omega} (\cdots)\: \L_\kappa(x,y)\: d\rho(y) \bigg)\, d\rho(x) \:,
\eeq
where $(\cdots)$ stands for a suitable
differential operator formed of jets.
A surface layer integral generalizes the concept of a surface integral over~$\partial \Omega$
to the setting of causal fermion systems.
The connection can be understood most easily in the
case when~$\L_\kappa(x,y)$ vanishes
unless~$x$ and~$y$ are close together. In this case, we only get a contribution to~\eqref{osi}
if both~$x$ and~$y$ are close to the boundary of~$\Omega$.
A more detailed explanation of the idea of a surface layer integral is given in~\cite[Section~2.3]{noether}.

We now recall those surface layer integrals for jets which will be of relevance in this paper.
\begin{Def} \label{defosi} We define the following surface layer integrals,
\begin{align*}
\sigma^\Omega \::\: \Jtest_0 \times \Jtest_0 &\rightarrow \R  \qquad \text{(symplectic form)} \notag \\
\sigma^\Omega(\u, \v) &= \int_{\Omega} d\rho(x) \int_{M \setminus \Omega} d\rho(y)\:
\big( \nabla_{1,\u} \nabla_{2,\v} - \nabla_{2,\u} \nabla_{1,\v} \big) \L(x,y) \\
(.,.)^\Omega \::\: \Jtest_0 \times \Jtest_0 &\rightarrow \R \qquad \text{(surface layer inner product)} \notag \\
(\u, \v)^\Omega &= \int_{\Omega} d\rho(x) \int_{M \setminus \Omega} d\rho(y)\:
\big( \nabla_{1,\u} \nabla_{1,\v} - \nabla_{2,\u} \nabla_{2,\v} \big) \L(x,y) \:. 
\end{align*}
\end{Def} \noindent
In order to make sure that the above surface layer integrals exist, we always assume that
the following regularity assumption holds (for details see~\cite[Section~3.5]{osi}):
\begin{Def} \label{defslr}
The jet space~$\Jtest$ is {\bf{surface layer regular}}
if~$\Jtest \subset \J^2$ and
if for all~$\u, \v \in \Jtest$ and all~$p \in \{1, 2\}$ the following conditions hold:
\begin{itemize}[leftmargin=2.5em]
\item[\rm{(i)}] The directional derivatives
\beq \nabla_{1,\u} \,\big( \nabla_{1,\v} + \nabla_{2,\v} \big)^{p-1} \L(x,y) \label{Lderiv1}
\eeq
exist.
\item[\rm{(ii)}] The functions in~\eqref{Lderiv1} are $\rho$-integrable
in the variable~$y$, giving rise to locally bounded functions in~$x$. More precisely,
these functions are in the space
\[ L^\infty_\text{\rm{loc}}\Big( L^1\big(M, d\rho(y) \big), d\rho(x) \Big) \:. \]
\item[\rm{(iii)}] The $\u$-derivative in~\eqref{Lderiv1} may be interchanged with the $y$-integration, i.e.
\[ \int_M \nabla_{1,\u} \,\big( \nabla_{1,\v} + \nabla_{2,\v} \big)^{p-1} \L(x,y)\: d\rho(y)
= \nabla_\u \int_M \big( \nabla_{1,\v} + \nabla_{2,\v} \big)^{p-1} \L(x,y)\: d\rho(y) \:. \]
\eitem
\end{Def}

\subsection{Weak Solutions in Lens-Shaped Regions} \label{secweak}
Similar to the procedure for hyperbolic partial differential equations, 
the initial value problem for the linearized field equations can be studied ``locally''
in lens-shaped regions as introduced in~\cite[Section~3]{linhyp}.
\begin{Def} \label{deflocfoliate}
Let~$U \subset M$ be an open subset of spacetime and~$I =[\tmin, t_{\max}]$ a compact interval.
Moreover, we let~$\eta \in C^\infty(I \times U, \R)$ be a function with~$0 \leq \eta \leq 1$ which for
all~$t \in I$ has the following properties:
\bitem
\item[{\rm{(i)}}] The function~$\theta(t,.) := \partial_t \eta(t,.)$ is non-negative and compactly supported in~$U$.
\item[{\rm{(ii)}}] For all~$x \in \supp \theta(t,.)$ and all~$y \in M \setminus U$,
the function~$\L(x,y)$ as well as its first and second derivatives in the direction of~$\Jtest_0$ vanish.
\eitem
We also write~$\eta(t,x)$ as~$\eta_t(x)$ and~$\theta(t,x)$ as~$\theta_t(x)$.
We refer to~$(\eta_t)_{t \in I}$ as a {\bf{local foliation}} inside~$U$.
\end{Def} \noindent

\begin{figure}
%
\psscalebox{1.0 1.0} 
{
\begin{pspicture}(-2.5,-1.9044089)(14.139044,1.9044089)
\definecolor{colour1}{rgb}{0.8,0.8,0.8}
\definecolor{colour0}{rgb}{0.6,0.6,0.6}
\definecolor{colour2}{rgb}{0.4,0.4,0.4}
\pspolygon[linecolor=colour1, linewidth=0.02, fillstyle=solid,fillcolor=colour1](1.1190436,-0.09187103)(1.7090436,-0.101871036)(2.3290436,-0.09187103)(3.1790435,-0.13187103)(3.8890436,-0.15187103)(4.4990435,-0.23187104)(5.1490436,-0.23187104)(5.7290435,-0.28187102)(5.3790436,-0.09187103)(4.8790436,0.23812896)(4.3590436,0.61812896)(3.7590437,0.92812896)(3.2690437,0.97812897)(2.7290435,0.81812894)(2.1090436,0.48812896)
\rput[bl](6.7190437,1.358129){\normalsize{$U \subset M := \supp \rho$}}
\psbezier[linecolor=black, linewidth=0.04](6.389044,1.298129)(5.585711,1.8936595)(2.0168202,2.0884824)(1.0690436,1.5881289672851562)(0.12126687,1.0877756)(-0.35428908,-0.8663405)(0.44904357,-1.461871)(1.2523762,-2.0574017)(5.331267,-1.9822243)(6.179044,-1.451871)(7.02682,-0.92151767)(7.192376,0.70259845)(6.389044,1.298129)
\psbezier[linecolor=colour0, linewidth=0.08](0.07904358,-0.15187103)(1.6590469,-0.024428569)(2.4390447,-0.15040691)(3.4390435,-0.15187103271484376)(4.4390426,-0.15333515)(5.51902,-0.4087202)(6.909044,-0.16187103)
\rput[bl](0.7290436,0.79812896){\normalsize{$\eta_t \equiv 0$}}
\rput[bl](0.8090436,-1.101871){\normalsize{$\eta_t \equiv 1$}}
\psbezier[linecolor=colour0, linewidth=0.08](0.7390436,-0.111871034)(1.4801531,-0.16686673)(2.449648,1.0228906)(3.4490435,0.9881289672851562)(4.448439,0.95336735)(5.2490635,-0.3373064)(5.9590435,-0.25187102)
\psbezier[linecolor=colour2, linewidth=0.08](1.0590435,-0.071871035)(1.870874,-0.09410611)(2.4199824,0.2114487)(3.4190435,0.16812896728515625)(4.4181046,0.124809235)(4.629186,-0.1793354)(5.7390437,-0.25187102)
\rput[bl](3.7290435,-0.93187106){\normalsize{$\supp \theta_t$}}
\psbezier[linecolor=black, linewidth=0.02, arrowsize=0.05291667cm 2.0,arrowlength=1.4,arrowinset=0.0]{->}(3.6090436,-0.791871)(3.0911489,-0.8228388)(3.0490437,-0.41154847)(2.8890436,0.16812896728515625)
\rput[bl](3.2790437,0.528129){\normalsize{$L$}}
\end{pspicture}
}
\caption{A local foliation.}
\label{figlocfol}
\end{figure}
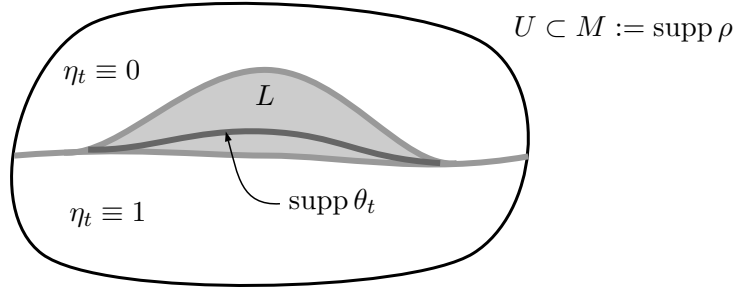

The parameter~$t$ can be thought of as the time of a local observer
and will often simply be referred to as {\em{time}}.
In the applications we always choose the functions~$\eta_t$
such that each function takes the values one and zero on non-empty subsets of~$U$
which can be thought of as the ``past'' and ``future'' of the support of the function~$\theta_t$
(see Figure~\ref{figlocfol}).
This support is referred to as the {\em{surface layer}} at time~$t$.
In order to avoid trivial cases, we always assume that~$\eta_t$ attains the values zero and one.
Then, by continuity, its image is the whole interval~$[0,1]$, i.e.\
\[ \eta_t(U) = [0,1] \qquad \text{for all~$t \in I$}\:. \]
It is the region of spacetime described by the local foliation.
The condition~(i) implies that the set~$L$ defined by
\beq \label{Ldef}
L := \bigcup_{t \in I} \supp \theta_t
\eeq
is compact. Since~$L$ is compact, we may restrict attention to the case that~$U$ is {\em{relatively
compact}}. This will always be assumed from now on.

For the following constructions, it will be useful to combine the
functions~$\eta_t$ and~$\theta_t$ with the measure~$\rho$ to form new measures:
The measure
\[ 
d\rho_t(x) := \theta_t(x)\: d\rho(x) \label{rhot} \]
with~$t \in I$ is supported in the surface layer at time~$t$.
For the measures supported in
a spacetime strip, we use the notation
\[ 
\eta_{[t_0, t_1]}\: d\rho \qquad \text{with} \qquad 
\eta_{[t_0,t_1]} := \eta_{t_1} - \eta_{t_0} \in C^\infty_0(U) \:, \]
where we always choose~$t_0,t_1 \in I$ with~$t_0 \leq t_1$.

In order to make the following constructions work, we need to restrict attention to jets in
a suitably chosen subspace of~$\Jtest$ denoted by
\[ 
\Jvary \subset \Jtest \:. \]
This jet space can be chosen arbitrarily, subject to conditions to be specified below.
Similar to~\eqref{J0def}, $\Jvary_0$ denotes the compactly supported jets in~$\Jvary$.
We now introduce softened versions of the surface layer integrals in Definition~\ref{defosi}
obtained by replacing the characteristic function by the cutoff functions~$\eta_t$
and~$(1-\eta_t)$. Moreover, we restrict these bilinear forms to~$\Jvary_0$,
\begin{align*}
&\sigma^t \::\: \Jvary_0 \times \Jvary_0 \rightarrow \R  \qquad \text{(softened symplectic form)} \notag \\
&\;\;\sigma^t(\u,\v) = \int_U d\rho(x)\: \eta_t(x) \int_U d\rho(y)\: \big(1-\eta_t(y)\big)
\: \Big( \nabla_{1,\u} \nabla_{2,\v} - \nabla_{2,\u} \nabla_{1,\v} \Big) \L(x,y) \\
&(.,.)^t \::\: \Jvary_0 \times \Jvary_0 \rightarrow \R \qquad \text{(softened surface layer inner product)} \notag \\
&\;\;(\u, \v)^t = \int_U d\rho(x)\: \eta_t(x) \int_U d\rho(y)\: \big(1-\eta_t(y)\big)
\: \Big( \nabla_{1,\u} \nabla_{1,\v} - \nabla_{2,\u} \nabla_{2,\v} \Big) \L(x,y) \:. 
\end{align*}
The main advantage of the softening is that it becomes possible to differentiate with respect to the
time parameter. The key observation is that the surface layer inner product satisfies
for any~$\v \in \J^\vary_0$ the {\em{energy identity}}
\beq \label{enid}
\frac{d}{dt}\: (\v, \v)^t 
= 2 \int_U \la \v, \Delta \v \ra(x) \: d\rho_t(x) -2 \int_U \Delta_2[\v, \v] \: d\rho_t(x)
+ \s \int_U b(x)^2\:d\rho_t(x)
\eeq
with
\[ 
\Delta_2[\v, \v] \big(x)
:= \frac{1}{2} \int_{M} \big( \nabla_{1, \v} + \nabla_{2, \v} \big)^2 \L(x,y)\: d\rho(y)\
- \frac{1}{2}\: \nabla_{\v}^2 \,\s\:. \]
Here we again use the convention that the ``partial jet derivatives'' do not act on
jets contained in other derivatives; in particular,
\[ \big( \nabla_{\v}^2 \, \s\big)(x) = b(x)^2\, \s \:, \]
where~$b$ denotes the scalar component of~$\v$.

In order to make use of the energy identity~\eqref{enid} for energy estimates,
we need to impose so-called hyperbolicity conditions, which we now motivate and introduce.
The starting point is the observation that for systems in Minkowski space~\cite{action},
the surface layer inner product~$(.,.)^t$ is positive definite for physically interesting jets,
provided that~$\eta_t$ is equal to one in the past and equal to zero in the future
of a spacelike hyperplane. With this in mind, it is sensible to assume that~$(\v,\v)^t$ is positive.
The lower bound in~\eqref{hypcond} is a stronger and more quantitative version of positivity:
\begin{Def} \label{defhypcond}
The local foliation~$(\eta_t)_{t \in I}$ inside~$U$
satisfies the {\bf{hyperbolicity condition}}
if there is a constant~$C>0$ such that for all~$t \in I$,
\beq
(\v, \v)^t \geq \frac{1}{C^2} \int_U \Big( \|\v(x)\|_x^2\: + \big|\Delta_2[\v, \v]\big| \Big) \: d\rho_t(x) 
\qquad \text{for all~$\v \in \Jvary_0$} \:, \label{hypcond}
\eeq
where~$\| . \|_x$ denotes the norm corresponding to the scalar product~\eqref{vsprod}.
\end{Def} \noindent
These hyperbolicity conditions also pose constraints for the choice of the functions~$\eta_t$;
these constraints can be understood as replacing the condition in the theory of hyperbolic PDEs
that the initial data surface be spacelike.
In general situations, the inequality~\eqref{hypcond}
is not obvious and must be arranged and verified in the applications.
More specifically, one can use the freedom in choosing the jet spaces~$\Jtest$ and~$\Jvary$,
the Riemannian metric in the scalar product~\eqref{vsprod} and the functions~$\eta_t$
in order to ensure that~\eqref{hypcond} holds.
Clearly, the smaller the jet space~$\Jvary$ is chosen, the easier it is to satisfy~\eqref{hypcond}.
The drawback is that the Cauchy problem will be solved in a weaker sense.

\begin{Def} \label{deflens}
A compact set~$L \subset M$ is a {\bf{lens-shaped region}} inside~$U$
if there is a local foliation~$(\eta_t)_{t \in I}$ inside~$U$
satisfying~\eqref{Ldef} which satisfies the {\bf{hyperbolicity conditions}}
of Definition~\ref{defhypcond}.
\end{Def}

In preparation of setting up the initial value problem,
we need to specify what we mean by ``$\v$ vanishes in the past of~$\tmin$.''
It is most convenient to implement all the necessary conditions in the definition of the jet space
\begin{align*} 
\underline{\J}_L := \big\{ \u \in \Jvary_0 \:\big|\: \eta_{t_{\min}}\, \u &\equiv 0 \quad \text{and} \\
(\u,\v)^{\tmin}&=0=\sigma^\tmin(\u,\v) \text{ for all~$\v \in \Jvary_0$} \big\} \:.
\end{align*}
Similarly, we define the space of jets which vanish in the future of~$\tmax$ by
\begin{align*} 
\overline{\J}_L := \big\{ \u \in \Jvary_0 \:\big|\: \big(1-\eta_{t_{\max}}\big)\, \u &\equiv 0 \quad \text{and} \\ \quad
(\u,\v)^{\tmax} &=0=\sigma^\tmax(\u,\v) \text{ for all~$\v \in \Jvary_0$} \big\} \:.
\end{align*}
Weak solutions of the Cauchy problem are defined as follows.
\begin{Def} \label{defweak}
A jet~$\v \in L^2(L)$ is a {\bf{weak solution}} of the Cauchy problem with zero initial data if
\beq \label{weak}
\la \Delta \u, \v \ra_{L^2(L)} = \la \u, \w \ra_{L^2(L)} \qquad \text{for all~$\u \in \overline{\J}_L$}\:.
\eeq
\end{Def}
The existence of weak solutions was proved in~\cite[Theorem~3.15]{linhyp}, inspired by
energy methods invented by K.O.\ Friedrichs for
symmetric hyperbolic systems in~\cite{friedrichs}; see also~\cite[Section~5.3]{john}
and~\cite[Chapter~11]{intro}.

\begin{Thm} \label{thmexist}  {\bf{(existence)}} Assume that~$L$ is a lens-shaped region inside~$U$
with foliation~$(\eta_t)_{t \in I}$ with~$I=[\tmin,\tmax]$. 
Then for every~$\w \in L^2(L)$ there is a weak solution~$\v \in L^2(L)$
of the Cauchy problem~\eqref{weak}. This solution is bounded by
\beq \label{vbound}
\|\v\|_{L^2(L)} \leq \Gamma\, \|\w\|_{L^2(L)} \:.
\eeq
\end{Thm} \noindent
We point out that, in general, weak solutions are not unique.
But the construction used for the proof of Theorem~\ref{thmexist} gives a
{\em{distinguished}} solution, which however depends on the choice of the lens-shaped region.

\section{Construction of Global Solutions} \label{secglobal}
We now come to the core of this paper: the construction of global solutions of the
linearized field equations.
We proceed in several steps. We begin with a weak solution in a lens-shaped region
as constructed in Theorem~\ref{weak}. By multiplying this solution with a cutoff function and extending
it by zero, we obtain a global weak solution. We show that the effect of the cutoff can be described with
an additional inhomogeneity~$\tilde{\w}$ supported in a neighborhood of the future boundary of the
lens-shaped region.
In the next step, we solve the linearized wave equation for the inhomogeneity~$-\tilde{\w}$,
giving rise to yet another inhomogeneity supported in the future of~$\tilde{\w}$.
Proceeding inductively and and adding all the solutions, we obtain the desired global solution.

\subsection{Local Solutions with Cutoff in the Future}
For technical simplicity, we assume that the Lagrangian has compact range
(for a variant of this definition and its usefulness we refer to~\cite[Definition~3.3]{noncompact}).
\begin{Def} \label{compactrange}
The Lagrangian is said to have {\bf{compact range}} on~$M$ if for any compact~$K\subset M$ there are a
compact set~${\mathfrak{K}}(K) \subset M$ as well as an open neighborhood~$\Omega \supset K$ such that
$$
\L(x,y)=0 \quad \text{if~$x \in \Omega$ and~$y\not\in {\mathfrak{K}}(K)$} \:.
$$
\end{Def} \noindent

\begin{figure}
\psscalebox{1.0 1.0} 
{
\begin{pspicture}(0,39.714012)(23.013618,66.24599)
\definecolor{colour0}{rgb}{0.8,0.8,0.8}
\definecolor{colour1}{rgb}{0.6,0.6,0.6}
\pspolygon[linecolor=colour0, linewidth=0.02, fillstyle=solid,fillcolor=colour0](17.24,40.925987)(18.29,40.405987)(19.02,40.035988)(20.08,39.775986)(20.88,39.89599)(21.49,40.295986)(22.08,40.605988)(22.52,40.775986)(22.19,40.955986)(21.67,41.295986)(21.15,41.675987)(20.55,41.98599)(20.06,42.035988)(19.52,41.875988)(18.89,41.635986)
\psbezier[linecolor=colour1, linewidth=0.08](16.99,40.91599)(17.900003,40.79343)(18.820002,39.897453)(19.94,39.77598754882813)(21.06,39.654522)(21.499977,40.48914)(22.71,40.815987)
\psbezier[linecolor=colour1, linewidth=0.08](16.98,40.93599)(17.72111,40.880993)(19.240604,42.08075)(20.24,42.04598754882812)(21.239395,42.011227)(22.04002,40.72055)(22.75,40.80599)
\rput[bl](18.89,40.66599){\normalsize{$L$}}
\psbezier[linecolor=black, linewidth=0.04, fillstyle=solid,fillcolor=colour1](19.73,40.267986)(19.754765,39.95148)(19.944963,39.960434)(20.16,39.995987548828126)(20.375036,40.03154)(20.649815,40.06234)(20.47,40.355988)(20.290186,40.649635)(19.705235,40.58449)(19.73,40.267986)
\rput[bl](19.91,40.135986){$W$}
\psbezier[linecolor=black, linewidth=0.04, fillstyle=solid,fillcolor=colour1](16.74,40.84599)(16.86491,40.54736)(17.723642,40.998783)(18.32,41.23598754882813)(18.916357,41.473194)(19.841461,41.97004)(20.7,41.625988)(21.558538,41.281937)(22.730513,40.408024)(22.95,40.635986)(23.169487,40.863953)(22.26957,41.342735)(21.55,41.73599)(20.83043,42.12924)(20.279997,42.253155)(19.93,42.21599)(19.580004,42.17882)(18.6958,41.950443)(18.41,41.82599)(18.124199,41.701534)(16.61509,41.144615)(16.74,40.84599)
\rput[bl](20.29,41.80599){$Z$}
\end{pspicture}
}
\caption{The sets~$W$ and~$Z$ of a lens-shaped region~$L$.}
\label{figWZ}
\end{figure}
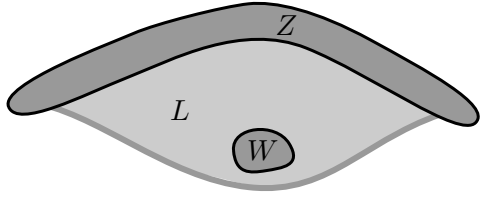

We let~$L$ be a lens-shaped regions inside~$U$,
with local foliation~$(\eta_t)_{t \in I}$ (see Definition~\ref{deflens}).
We define the compact set~$Z \subset M$ by
\beq
Z := \mathfrak{K}\Big( \overline{ \{x \in U \:|\: \eta_{t_{\max}}(x) <1 \} }^M \Big) 
\cap \mathfrak{K}\Big( \overline{ \{x \in U \:|\: \eta_{t_{\max}}(x) > 0 \} }^M \Big) 
\;\subset\; M \:.
\eeq
Moreover, we choose an open set
\[ W \subset \big\{ x \in {U} \:|\: {\eta}_{{I}}(x)=1 \big\}  \cap \big( L \setminus Z \big) \;\subset\; L \:. \]
The sets~$W$ and~$Z$ are illustrated in Figure~\ref{figWZ}.

We introduce a space of test jets which is compatible with multiplying by a cutoff function
in the sense that
\[ \J'_{L} = \{ \u \in \Jvary_0 \:|\: \eta_{t_{\max}} \u \in \overline{\J}_L \} \:. \]
Working with this modified test space complements the constructions in~\cite{linhyp, weyl} by a new idea:
Instead of changing the space~$\Jvary_0$, we suitably restrict the space of test jets
in such a way that they can be used to test the solution multiplied by a cutoff function.

\begin{Thm} \label{thmglue}
 For every~$\w \in L^2(W)$ there exists~$\tilde{\w} \in L^2(Z)$ together with a weak
solution~$\v \in L^2(M)$ of the inhomogeneous linearized field equations
\beq \label{weakLL}
\la \Delta \u , \v \ra_{L^2(M)} = \la \u, \w \ra_{L^2(M)} + \la \u, \tilde{\w} \ra_{L^2(M)} 
\qquad \text{for all~$\u \in \J'_{L}$}
\eeq
\end{Thm}
\Proof
Given~$\w \in \J_0({L})$, we let~${\v} \in L^2({L})$ be the corresponding weak solution, i.e.\
\beq \label{rhs}
\la \Delta \u, {\v} \ra_{L^2({L})} = \la \u, \w \ra_{L^2({L})} \qquad \text{for all~$\u \in \overline{\J}_L$}\:.
\eeq
The $L^2$-scalar product on the left involves the weight~${\eta}_{{I}}$, i.e.
\[ \la \Delta \u, {\v} \ra_{L^2({L})} = \int_{{L}} \la \Delta \u, {\v} \ra(x)\: {\eta}_{{I}}(x)\: d\rho(x) \:. \]
We absorb this weight into the weak solution,
\[ \la \Delta \u, {\v} \ra_{L^2({L})} = \int_{{L}} \la \Delta \u, \tilde{\v} \ra(x)\: d\rho(x) \qquad
\text{with} \qquad \tilde{\v}(x) := {\eta}_{{I}}(x)\: {\v}(x) \:. \]
By definition of~$W$, the function~${\eta}_{{I}}$ is identically equal to one on the support of~$\w$,
\beq \label{wvanish}
{\eta}_I\, \w = \w \:.
\eeq
Therefore, the weight function can be omitted on the right side of~\eqref{rhs}. 
Extending the jets~$\tilde{\v}$ and~$\w$ by zero to all of~$M$, we thus obtain the weak equation
\beq \label{weak2}
\la \Delta \u, \tilde{\v} \ra_{L^2(M)} = \la \u, \w \ra_{L^2(M)} \qquad \text{for all~$\u \in \overline{\J}_L$}\:.
\eeq

Next, we want to formulate a weak equation for the space of test jets
\[ \J'_{L} := \{ \u \in \Jvary_0 \:|\: {\eta}_{{t}_{\max}} \u \in \overline{\J}_L \} \:. \]
Thus let~$\u \in \J'_{L}$. Then~\eqref{weak2} holds for the test jet~${\eta}_{{t}_{\max}} \u$.
Using~\eqref{wvanish}, we obtain
\begin{align*}
\la \u, \w \ra_{L^2(M)} &= \la \u, {\eta}_{{t}_{\max}} \w \ra_{L^2(M)}
= \la \Delta \big( {\eta}_{{t}_{\max}} \u \big), \tilde{\v} \ra_{L^2(M)} \\
&= \la \Delta \u , \big( {\eta}_{{t}_{\max}} \tilde{\v} \big) \ra_{L^2(M)}
+ \la \big[ \Delta, {\eta}_{{t}_{\max}} \big] \u ,\tilde{\v} \ra_{L^2(M)} \:.
\end{align*}
We conclude that the jet~$\v := {\eta}_{{t}_{\max}} \tilde{\v}$ satisfies the weak equation~\eqref{weakLL}
with the additional inhomogeneity
\[ \tilde{\w} := \big[ \Delta, {\eta}_{{t}_{\max}} \big] \tilde{\v} \:. \]

Let us verify that~$\tilde{\w}$ is supported inside~$Z$. We first write out the commutator,
\begin{align}
\la \u, \tilde{\w}\ra(x) &= \la \u, \big[ \Delta, {\eta}_{{t}_{\max}} \big] \tilde{\v} \ra(x)
=\la \u, \Delta \big( {\eta}_{{t}_{\max}} \tilde{\v} \big) \ra(x) 
- \la {\eta}_{{t}_{\max}} \u, \Delta \tilde{\v} \ra(x) \notag \\
&= \nabla_\u \bigg( \int_M \big( \nabla_{1, {\eta}_{{t}_{\max}} \tilde{\v}} + \nabla_{2, {\eta}_{{t}_{\max}} \tilde{\v}} \big) \L(x,y)\: d\rho(y) - \nabla_{{\eta}_{{t}_{\max}} \tilde{\v}} \,\s \bigg) \notag \\
&\quad\: -{\eta}_{{t}_{\max}}(x)  \nabla_\u \bigg( \int_M \big( \nabla_{1,\tilde{\v}} + \nabla_{2, \tilde{\v}} \big) \L(x,y)\: d\rho(y) - \nabla_{\tilde{\v}} \,\s \bigg) \notag \\
&= \nabla_\u \int_M \big({\eta}_{{t}_{\max}}(y) - {\eta}_{{t}_{\max}}(x) \big)
\nabla_{2, \tilde{\v}} \L(x,y)\: d\rho(y) \:. \label{commutator}
\end{align}
Let
\[ x \not \in \mathfrak{K}\Big( \overline{ \{z \:|\: {\eta}_{{t}_{\max}}(z) <1 \} } \Big) \:. \]
Then~${\eta}_{{t}_{\max}}(x)=1$ and~$\nabla_{2, \tilde{\v}} \L(x,y)$ vanishes
unless~${\eta}_{{t}_{\max}}(y)=1$. Hence the integrand in~\eqref{commutator} vanishes identically.
The same argument applies if~$x \not \in \mathfrak{K}( \overline{ \{z \:|\: {\eta}_{{t}_{\max}}(z) >0 \} })$.
\QED

\subsection{Gluing Local Solutions} \label{secglue}
\begin{Def} \label{deflochyp} Spacetime~$M$ is called {\bf{locally hyperbolic}} if for every~$x \in M$ there is a lens-shaped region~$L \subset U$ with corresponding sets~$W \subset L$ open and~$Z \subset M$ compact as in Theorem~\ref{thmglue} with~$x \in W$.
\end{Def}
As a manifold, $\F$ is $\sigma$-compact. Therefore, we can cover a locally hyperbolic spacetime~$M$
by a sequence of sets~$W_\ell$, i.e.\
\[ M = \bigcup_{\ell \in \N} W_\ell \:. \]
For every~$W_\ell$, we denote the corresponding sets~$L$ and~$Z$ in Definition~\ref{deflochyp} by~$L_\ell$
and the~$Z_\ell$. We have the situation in mind that the sets~$W_\ell$ are chosen to be very small
compared to the size of the lens-shaped regions. 

\begin{figure}
\psscalebox{1.0 1.0} 
{
\begin{pspicture}(0,38.051514)(31.610802,61.647503)
\definecolor{colour0}{rgb}{0.8,0.8,0.8}
\definecolor{colour1}{rgb}{0.6,0.6,0.6}
\definecolor{colour2}{rgb}{0.4,0.4,0.4}
\pspolygon[linecolor=colour0, linewidth=0.02, fillstyle=solid,fillcolor=colour0](28.65887,40.180126)(29.194862,39.898815)(29.567503,39.69865)(30.108597,39.557995)(30.516972,39.622913)(30.828356,39.839306)(31.129532,40.00701)(31.354137,40.098976)(31.185682,40.196354)(30.92024,40.38029)(30.654797,40.585865)(30.348516,40.753567)(30.098389,40.780617)(29.822737,40.69406)(29.501143,40.564224)
\psbezier[linecolor=colour1, linewidth=0.04](28.531254,40.174717)(28.99578,40.108414)(29.46541,39.623703)(30.037132,39.55799438476563)(30.608854,39.492283)(30.833448,39.943798)(31.451126,40.120617)
\pspolygon[linecolor=colour0, linewidth=0.02, fillstyle=solid,fillcolor=colour0](27.538872,39.540127)(28.074862,39.258816)(28.447502,39.05865)(28.988598,38.917995)(29.39697,38.982914)(29.708355,39.199306)(30.009531,39.367012)(30.344137,39.51898)(30.145683,39.636356)(29.82024,39.88029)(29.554796,40.045864)(29.228518,40.113567)(28.978388,40.140617)(28.702736,40.054058)(28.381142,39.924225)
\psbezier[linecolor=colour1, linewidth=0.04](27.411255,39.534718)(27.875782,39.468414)(28.34541,38.983704)(28.917131,38.91799438476563)(29.488855,38.852283)(29.71345,39.3038)(30.331125,39.480618)
\pspolygon[linecolor=colour0, linewidth=0.02, fillstyle=solid,fillcolor=colour0](26.868872,38.710125)(27.404861,38.428814)(27.777502,38.22865)(28.318598,38.087994)(28.72697,38.152912)(29.038357,38.369305)(29.33953,38.53701)(29.564137,38.62898)(29.395683,38.726357)(29.13024,38.91029)(28.864798,39.115864)(28.558517,39.28357)(28.308388,39.310616)(28.032736,39.22406)(27.711142,39.094223)
\psbezier[linecolor=colour1, linewidth=0.04](26.741255,38.704716)(27.205782,38.638412)(27.67541,38.153706)(28.247131,38.08799438476562)(28.818855,38.022285)(29.04345,38.473797)(29.661125,38.650616)
\psbezier[linecolor=colour1, linewidth=0.08](26.736149,38.715534)(27.114462,38.685783)(27.890114,39.33483)(28.400272,39.31602717165086)(28.91043,39.297222)(29.319122,38.598988)(29.681543,38.645206)
\psbezier[linecolor=black, linewidth=0.02, fillstyle=solid,fillcolor=colour2](28.139935,38.354156)(28.152576,38.182934)(28.249666,38.18778)(28.359434,38.20701077820828)(28.469204,38.226246)(28.609468,38.24291)(28.51768,38.401764)(28.42589,38.560623)(28.127293,38.525383)(28.139935,38.354156)
\psbezier[linecolor=black, linewidth=0.02, fillstyle=solid,fillcolor=colour1](26.613638,38.666847)(26.6774,38.505295)(27.115755,38.749508)(27.420176,38.87783045033937)(27.724596,39.006153)(28.19683,39.27494)(28.635086,39.088814)(29.073343,38.902687)(29.671597,38.429916)(29.783638,38.55324)(29.895678,38.676567)(29.4363,38.93558)(29.068983,39.148323)(28.701668,39.361065)(28.420689,39.4281)(28.242027,39.407993)(28.063366,39.387886)(27.61201,39.26434)(27.466118,39.19701)(27.320225,39.129684)(26.549875,38.8284)(26.613638,38.666847)
\psbezier[linecolor=colour1, linewidth=0.08](27.406149,39.545536)(27.784462,39.515785)(28.560114,40.164833)(29.070272,40.14602717165086)(29.580431,40.12722)(30.149122,39.438988)(30.511545,39.485207)
\psbezier[linecolor=black, linewidth=0.02, fillstyle=solid,fillcolor=colour2](28.809935,39.18416)(28.822577,39.012936)(28.919666,39.017776)(29.029434,39.03701077820828)(29.139204,39.056244)(29.279469,39.072906)(29.18768,39.231766)(29.09589,39.39062)(28.797293,39.35538)(28.809935,39.18416)
\psbezier[linecolor=black, linewidth=0.02, fillstyle=solid,fillcolor=colour1](27.283638,39.496845)(27.3474,39.335293)(27.785755,39.579506)(28.090176,39.70783045033937)(28.394596,39.836155)(28.876831,40.19494)(29.315086,40.008816)(29.753342,39.82269)(30.451597,39.319916)(30.563637,39.44324)(30.675678,39.566566)(30.136301,39.86558)(29.768984,40.078323)(29.401667,40.291065)(29.090689,40.258102)(28.912027,40.237995)(28.733366,40.217888)(28.28201,40.094337)(28.136118,40.02701)(27.990225,39.959682)(27.219875,39.658398)(27.283638,39.496845)
\psbezier[linecolor=colour1, linewidth=0.08](28.52615,40.185535)(28.90446,40.155785)(29.680113,40.804832)(30.190271,40.78602717165086)(30.70043,40.767223)(31.109123,40.06899)(31.471544,40.115208)
\psbezier[linecolor=black, linewidth=0.02, fillstyle=solid,fillcolor=colour2](29.759933,39.77416)(29.772575,39.602936)(29.869665,39.607777)(29.979435,39.62701077820828)(30.089205,39.646244)(30.349468,39.65291)(30.13768,39.821766)(29.92589,39.990623)(29.747292,39.94538)(29.759933,39.77416)
\psbezier[linecolor=black, linewidth=0.02, fillstyle=solid,fillcolor=colour1](28.403637,40.13685)(28.4674,39.975296)(28.905756,40.219505)(29.210176,40.34783045033937)(29.514597,40.476154)(29.986832,40.74494)(30.425087,40.558815)(30.863342,40.37269)(31.461597,39.899914)(31.573637,40.02324)(31.685678,40.146564)(31.226301,40.40558)(30.858984,40.61832)(30.491667,40.831066)(30.21069,40.8981)(30.032028,40.877995)(29.853367,40.857887)(29.40201,40.73434)(29.256117,40.66701)(29.110226,40.59968)(28.339876,40.2984)(28.403637,40.13685)
\rput[bl](27.45,38.4){$L_{\ell_1}$}
\rput[bl](28.4,38.46){$W_{\ell_1}$}
\rput[bl](29.1,39.25){$W_{\ell_2}$}
\rput[bl](30.0,39.9){$W_{\ell_3}$}
\end{pspicture}
}
\caption{Future-related sets.}
\label{figfuturerel}
\end{figure}
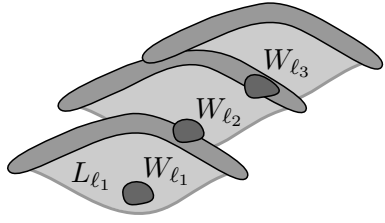

\begin{Def} The set~$W_{\ell'}$ is {\bf{future-related}} to~$W_\ell$ if there is a
finite sequence
\[ \ell = \ell_1, \ell_2, \ldots \ell_p = \ell' \in \N \]
such that
\[ W_{\ell_{i+1}} \cap Z_{\ell_i} \neq \varnothing \qquad \text{for all~$i=1, \ldots, p-1$}\:. \]
\end{Def} \noindent
This notion is illustrated in Figure~\ref{figfuturerel}.

\begin{Def} \label{defstronglycausal}
Spacetime~$M$ is {\bf{strongly causal}} if there is a covering~$(W_\ell)_{\ell \in \N}$
such that the following implication holds,
\beq \label{stableimpl}
\text{$W_{\ell'}$ is future-related to~$W_{\ell}$} \qquad \Longrightarrow \qquad W_\ell \cap W_{\ell'} = \varnothing\:.
\eeq
\end{Def}

From now on, we assume that~$M$ is strongly causal. We let~$(W_\ell)_{\ell \in \N}$ be a covering
having the property~\eqref{stableimpl}.
We introduce the test space
\[ \J' := \bigcap_{\ell \in \N} \J'_{L_\ell} \:. \]

\begin{Thm} \label{thmglobal}
Assume that the Lagrangian has compact range (see Definition~\ref{compactrange})
and that spacetime is strongly causal with chosen covering~$(W_\ell)_{\ell \in \N}$
(see Definition~\ref{defstronglycausal}).
Then for any~$\w \in L^2_0(M)$ there is a
global weak solution~$\v \in L^2_\loc(M)$,
\[ 
\la \Delta \u, \v \ra_M = \la \u, \w \ra_{L^2(M)} \qquad \text{for all~$\u \in \J'$}\:. \]
\end{Thm}
\Proof We proceed inductively. In the first step, we choose~$p$ such that the
support of~$\w$ lies in the union of the sets~$W_1, \ldots, W_p$. For each inhomogeneity
\[ \w_k := \chi_{W_k \setminus (W_1 \cup \cdots \cup W_{k-1})} \w \]
we let~$\v_k$ be the corresponding distinguished solution constructed in Theorem~\ref{thmglue},
and we let~$\tilde{\w}_k \in L^2(Z_k)$ be the additional inhomogeneity.

In the next step, we cover each~$Z_k$ by a finite number of the sets~$W_1, \ldots, W_{p'}$,
choose the inhomogeneities
\[ -\chi_{W_\ell \setminus (W_1, \ldots, W_{\ell-1})} \tilde{\w}_k \]
and apply Theorem~\ref{thmglue} to each of these inhomogeneities.
Adding up all the solutions, the additional inhomogeneities~$\tilde{\w}_k \in L^2(Z_{\ell_k})$
obtained in the first step have been compensated.
But we have generated new inhomogeneities, which can be compensated in the third step.
We iterate this procedure. Adding all the solutions, we get a formal weak solution to the inhomogeneous
linearized field equations with inhomogeneity~$\w$.

In order to prove that the series converges, we prove that the sum over all the solutions is locally
finite. To this end, let~$K \subset M$ be compact. We choose~$N$ such that
\[ K \subset W_1, \ldots, W_N \:. \]
It follows from our inductive construction that the sets~$W_\ell$ lie in the future of the
corresponding sets in the previous step. Therefore, the assumption~\eqref{stableimpl}
implies that only a finite number of these sets intersect~$K$.
\QED

Before going on, we point out that the global weak solution constructed in this theorem
depends on the choice of the sequence~$(L_\ell)_{\ell \in \N}$ of lens-shaped regions
(note that reordering the sequence might give a different weak solution).
This dependence cannot be avoided in view of the intrinsic non-uniqueness of weak solutions
mentioned in the introduction. But, having chosen such a sequence, Theorem~\ref{thmglobal}
gives a distinguished solution with nice support properties.

\section{Construction of Causal Green's Operators} \label{segreen}
Having chosen a sequence~$(L_\ell)_{\ell \in \N}$ of lens-shaped regions,
we introduce the {\em{retarded Green's operator}} as the
uniquely defined mapping
\[ S^\wedge \::\: L^2_0(M) \rightarrow L^2_\loc(M)\:,\qquad \w \mapsto -\v \:. \]
Similarly, solving the Cauchy problem to the past, we obtain the {\em{advanced Green's operator}}
\[ S^\vee \::\: L^2_0(M) \rightarrow L^2_\loc(M)\:. \]
The difference of the causal Green's operators is the {\em{causal fundamental solution}}
\[ G := S^\vee - S^\wedge \::\: L^2_0(M) \rightarrow L^2_\loc(M) \:. \]
It maps to global weak solutions in the sense that for all~$w \in L^2_0(M)$,
\[ \la \Delta \u, G \w \ra_M = 0 \qquad \text{for all~$\u \in \J'$}\:. \]
We remark that, using the energy estimates~\eqref{vbound}, one sees that the operator~$G$
is a continuous mapping from~$L^2_0(M)$ to~$L^2_\loc(M)$ (with respect to the
corresponding topologies). If~$M$ is assumed to have a smooth manifold structure,
one can apply the Schwartz kernel theorem~\cite[Theorem~5.2.1]{hormanderI}
to represent the Green's operator by an integral operator whose kernel is a
bi-distribution in~$\D'(M \times M)$.

In the above definitions of~$S^\wedge$, $S^\vee$ and~$G$,
we chose the domain of definition as large as possible.
For the applications, however, it is convenient to restrict attention to a smaller domain of ``nice'' jets.
In this way, the operators~$\Delta$ and~$G$ should have the properties which are summarized
in the exact sequence~\eqref{exact}.
The following constructions are similar as in~\cite[Section~5.3]{linhyp}, but with two major differences.
First, for technical simplicity, for the domain of the Green's operators and the causal fundamental solution,
we restrict attention to compactly supported jets (the generalization to jets vanishing in the past or future
could be carried out similar as explained in~\cite[Remark~5.8]{linhyp}).
Second, we prefer to define the jet spaces without taking equivalence classes.
More precisely, we define
\begin{align*}
\J^{**}_0 &:= \big\{ \u \in \Jvary_0 \:\big|\: S^\vee \Delta \u = S^\wedge \Delta \u = -\u \big\} \\
\J^*_0 &:= \big\{ \u \in L^2_0(M, d\rho) \:\big|\: S^\vee \u, S^\wedge \u \in \Jvary 
\text{ and } \Delta S^\vee \u = \Delta S^\wedge \u = -\u \big\}  \\
\J_\sc &:= \big\{ S^\wedge \u_1 + S^\vee \u_2 \:\big|\: \u_1, \u_2 \in L^2_{0}(M, d\rho)\:,\;\;
S^\wedge \u_1, S^\vee \u_2 \in \Jvary, \\
&\qquad\qquad\qquad\qquad\qquad\qquad \: \Delta S^\wedge \u_1 = -\u_1 \,\text{ and } \Delta S^\vee \u_2 = -\u_2 \big\} \\
\J^*_\sc &:= \big\{ \u_1 + \u_2 \:\big|\: \u_1, \u_2 \in L^2_{0}(M, d\rho)\:,\;\;
S^\wedge \u_1, S^\vee \u_2 \in \Jvary, \\
&\qquad\qquad\qquad\qquad\qquad\qquad \: \Delta S^\wedge \u_1 = -\u_1 \,\text{ and } \Delta S^\vee \u_2 = -\u_2 \big\} \:.\end{align*}
In the next theorem we combine the properties of the causal fundamental solution
in an exact sequence, similar as obtained for linear hyperbolic PDEs in
globally hyperbolic space-times in~\cite[Proposition~8]{ginouxML}
and~\cite[Theorem~4.3]{baergreen}.

\begin{Thm} \label{thmexact}
Assume that the Lagrangian has compact range (see Definition~\ref{compactrange})
and that spacetime is strongly causal with chosen covering~$(W_\ell)_{\ell \in \N}$
(see Definition~\ref{defstronglycausal}).
Then the sequence~\eqref{exact} is exact.
\end{Thm}
\Proof We proceed in several steps:
\begin{itemize}[leftmargin=2.5em]
\item[(i)] $\Delta (\J_0^{**}) \subset \J_0^*$: This follows immediately from the definitions, noting that
for any~$\u \in \J_0^{**}$,
\[ \Delta S^\vee (\Delta u) = \Delta \big( S^\vee \Delta u) = \Delta \big( - u) = - \Delta u \:, \]
and similarly for~$S^\wedge$.
\item[(ii)] $\Delta : \J^{**}_0 \rightarrow \J^*_0$ is injective: 
Let~$\v \in \J^{**}_0$ with~$\Delta \v=0$. 
Multiplying by~$S^\vee$
and using again the definition of~$\J^{**}_0$, we conclude that~$\v = -S^\vee \Delta \v = 0$.
\item[(iii)] If~$G \u = 0$ for~$\u \in \J^*_0$, then~$\u$ can be represented
as~$\u =\Delta \v$ with~$\v \in \J^{**}_0$:
By definition of~$G$ and~$\J^*_0$, we know that
\[ \v:= -S^\vee \u = -S^\wedge \u \in \Jvary \:. \]
Being supported both in the causal future and in the past, it follows that~$\v$ is compactly supported.
Finally, the equation~$\Delta \v = \u$ follows by definition of~$\J_0^*$.
\item[(iv)] $G( \J_0^*) \subset \J_\sc$: Follows immediately by choosing~$\u_1=\u_2=u$.
\item[(v)] The product~$G \circ \Delta : \J^{**}_0 \rightarrow \J_\sc$ vanishes: Let~$\v \in \J^{**}_0$.
Then, by definition of~$G$ and~$\J^{**}_0$,
\[ G \Delta \v = S^\wedge \Delta \v - S^\vee \Delta \v = -\v + \v = 0 \:. \]
\item[(vi)] If~$\Delta \v = 0$ for~$\v \in \J_\sc$, then~$\v$ can be represented
as~$\v =G \u$ with~$\u \in \J^*_0$: Representing~$\v$ as in the definition of~$\J_\sc$, we obtain
by definition of~$\J^*_0$ that~$0=-\Delta \v = \u_1 + \u_2$.
Therefore~$\v = G \u$ with~$\u:=\u_2=-\u_1$. Moreover, one readily verifies that~$\u \in \J^*_0$.
\item[(vii)] $\Delta(\J_\sc) \subset \J^*_\sc$: This follows immediately from the definitions.
\item[(viii)] The product~$\Delta \circ G : \J_0^* \rightarrow \J_\sc^*$ vanishes:
This follows immediately from the definitions of the jet spaces.
\item[(ix)] The operator~$\Delta : \J_\sc \rightarrow \J^*_\sc$ is surjective:
Let~$\u \in \J^*_\sc$. According to the definition of~$\J^*_\sc$, we can represent~$\u$ as
\[ \u = \u_1 + \u_2 \qquad \text{with~$\u_1 \in L^2_{0}(M, d\rho)$ and~$\u_2 \in L^2_{0}(M, d\rho)$}\:. \]
Then by definition, the jet~$\v := -S^\wedge \u_1 - S^\vee \u_2$ is in~$\J_\sc$
and~$\Delta \v = \u$.
\eitem
This concludes the proof.
\QED

The image of the operator~$G$ in the exact sequence~\eqref{exact} are the linearized weak
solutions of spatially compact support denoted by
\[ 
\Jlin_\sc := G  \,\J^*_0  \subset \Jtest \:. \]

\section{Causal Cone Structures} \label{seccones}
We define the causal future of an open set~$V \subset M$ by
\[ J^\vee(V) = \bigcup_{\w \in L^2_0(V)} \big\{ \supp \v \:\big|\: \text{$\v$ global weak solution} \big\} \:, \]
with the global weak solution as constructed in Theorem~\ref{thmglobal}.
We say that~$y$ lies in the {\em{future}} of~$x$ if~$y \in J^\vee(V)$ for every open neighborhood~$V$ of~$x$.
We consider the relation
\[ \hat{R} := \{ (x,y) \in M \times M \:|\: \text{$y$ lies in the future of~$x$} \} \subset M \times M \:. \]
We now take the transitive closure of~$R$ defined by
\[ R := \bigcap \big\{ S \subset M \times M \:\big|\: S \supset \hat{R} \text{ is transitive and closed} \big\} \:. \]
For pairs~$(x,y) \in R$ we say that~$y$ lies in the {\em{transitive causal future}} of~$x$.
We denote the points in the transitive causal future of~$x$ by~$J^\vee_x$.

Taking our constructions as a starting point, one can also introduce open light cones and 
other related structures. Our setting is more general than that of differential geometry
because the spacetime~$M$ could be singular or discrete. Moreover, we do not need to make
any smoothness assumptions. Working out the connection to the causal structures
in differential geometry as described in~\cite{minguzzi+sanchez} is an important problem of
future research. Moreover, in the case that~$M$ has a smooth manifold structure, it would be
desirable to associate the causal structure to closed cone structures in the tangent bundle,
with the goal of making the methods and results for globally hyperbolic closed cone fields
(see for example~\cite{minguzzi-cones, bernard-suhr}) applicable to
causal variational principles.

\Thanks {{\em{Acknowledgments:}} We would like to thank Sami Abdallah, Claudio Dappiaggi and
Miguel S{\'a}nchez Caja for helpful discussions. We are grateful to the referees for valuable suggestions.

\providecommand{\bysame}{\leavevmode\hbox to3em{\hrulefill}\thinspace}
\providecommand{\MR}{\relax\ifhmode\unskip\space\fi MR }
\providecommand{\MRhref}[2]{%
  \href{http://www.ams.org/mathscinet-getitem?mr=#1}{#2}
}
\providecommand{\href}[2]{#2}

\end{document}